\documentclass[conference]{IEEEtran}
\usepackage[letterpaper, top=1.91cm, bottom=2.45cm, left=1.62cm, right=1.62cm]{geometry}
\usepackage{amsmath,amsfonts}
\usepackage{array}
\usepackage[caption=false,font=footnotesize,labelfont=rm,textfont=rm]{subfig}
\usepackage{textcomp}
\usepackage{stfloats}
\usepackage{url} 
\usepackage{verbatim}
\usepackage{graphicx}
\usepackage{cite}
\usepackage{url}
\usepackage{amsmath}
\usepackage{amsthm}
\usepackage{amssymb}
\usepackage{upgreek}
\usepackage{bm}
\usepackage[utf8]{inputenc}
\usepackage{diagbox} 
\usepackage{xcolor}
\usepackage{multirow}
\usepackage{booktabs}
\usepackage{tabularx}
\usepackage{cleveref}
\usepackage{url}
\usepackage{amsmath}
\usepackage{amsthm}
\usepackage{amssymb}
\usepackage{upgreek}
\usepackage{cite}
\usepackage{mathtools}
\usepackage{color}
\usepackage{subfig}
\usepackage{bm}

\usepackage[ruled, vlined, linesnumbered]{algorithm2e}
\SetKwInput{KwInit}{Init}
\SetKwInOut{KwRequire}{Require}
\SetKwInOut{KwEnsure}{Ensure}
\SetKwComment{Comment}{// }{}
\SetCommentSty{mycommfont}
\newcommand{\Break}{$\mathbf{break}$}

\SetCommentSty{mycommfont}

\begin{document}
\title{Joint Optimization of Offloading, Batching and DVFS for Multiuser Co-Inference}  
\author{
    \IEEEauthorblockN{Yaodan Xu, Sheng Zhou, Zhisheng Niu}
    \IEEEauthorblockA{Department of Electronic Engineering, Tsinghua University, Beijing 100084, P.R. China}
    
    \IEEEauthorblockA{xyd21@mails.tsinghua.edu.cn, \{sheng.zhou, niuzhs\}@tsinghua.edu.cn}
}
\maketitle

\begin{abstract}
With the growing integration of artificial intelligence in mobile applications, a substantial number of deep neural network (DNN) inference requests are generated daily by mobile devices.
Serving these requests presents significant challenges due to limited device resources and strict latency requirements.
Therefore, edge-device co-inference has emerged as an effective paradigm to address these issues. In this study, we focus on a scenario where multiple mobile devices offload inference tasks to an edge server equipped with a graphics processing unit (GPU). For finer control over offloading and scheduling, inference tasks are partitioned into smaller sub-tasks. Additionally, GPU batch processing is employed to boost throughput and improve energy efficiency. 
This work investigates the problem of minimizing total energy consumption while meeting hard latency constraints.
We propose a low-complexity Joint DVFS, Offloading, and Batching strategy (J-DOB) to solve this problem.
The effectiveness of the proposed algorithm is validated through extensive experiments across varying user numbers and deadline constraints. 
Results show that J-DOB can reduce energy consumption by up to 51.30\% and 45.27\% under identical and different deadlines, respectively, compared to local computing.
\end{abstract}

\section{Introduction}
Recent advancements in deep learning have driven the widespread adoption of deep neural networks (DNNs) in mobile applications. 
However, these applications often come with strict latency requirements, and achieving reliable inference may involve large DNN models, which can require substantial computational and energy resources.
As a result, performing fast and reliable inference at the resource-constrained network edge poses significant challenges.
To address this, we consider scenarios where mobile devices can partition DNN inference tasks into smaller sub-tasks and offload portions to an edge server—a process known as edge-device co-inference. 
Equipped with more powerful computing resources, the edge server can efficiently accelerate the processing of DNN tasks.


Graphics processing units (GPUs), neural-network processing units (NPUs), and other custom-designed deep neural network (DNN) hardware accelerators have become essential for accelerating DNN computations, mainly because of their outstanding parallel processing capabilities. Batch processing, commonly referred to as \emph{batching}, is a vital technique that enhances the utilization of  parallelism by aggregating multiple tasks into a single batch for concurrent processing. 
The number of tasks collected in a batch is called the \emph{batch size}.
This approach significantly improves both throughput and energy efficiency during inference \cite{choi2021lazy} and training \cite{you2019large}.
However, existing studies on edge-device co-inference \cite{hu2019dynamic,song2021adaptive,tang2020joint} cannot be directly applied to these DNN hardware accelerators. The reason lies in the distinct differences in computing resource allocation and task scheduling necessitated by batch processing. 

The issue of  inference task batching has been investigated in  inference serving systems \cite{zhang2019mark,xu2023smdp,ali2020batch,choi2021lazy,10829781}.
While larger batches can enhance throughput and energy efficiency, they may also lead to increased latency due to the need to wait for sufficient tasks and the extended inference time required to process multiple requests at once. Consequently, optimizing batch size is essential for achieving a better balance between responsiveness and energy efficiency. Implementing an effective dynamic batching policy can significantly lower power consumption \cite{10829781}.
However, existing approaches are not directly applicable to device-edge co-inference scenarios because they typically offload the entire inference task to the server and overlook offloading latency. 

Several recent works have jointly considered multi-user offloading and batch scheduling\cite{shi2022multiuser,liu2023resource,cang2024joint}.
The authors of \cite{liu2023resource} propose an adaptive strategy to select tasks to be offloaded and batch processed at the edge, incorporating bandwidth allocation and early exit mechanisms.
Another work \cite{cang2024joint} aims to maximize the throughput by designing a dynamic batch scheduling scheme.
Nevertheless, both of them consider entire task offloading rather than the fine-grained co-inference scenario.
Our work builds upon \cite{shi2022multiuser}, which considers the device energy consumption minimization under hard latency constraints.
The complexity of this problem is heightened by the time constraints introduced by batching and the selection of offloading parts for co-inference. 
A complete optimization problem of offloading and batching is formulated in \cite{shi2022multiuser} and claimed to be NP-hard.
To address this, \cite{shi2022multiuser} 
proposes a heuristic algorithm based on assumptions of size-independent batch processing time and identical deadlines.

Dynamic Voltage and Frequency Scaling (DVFS) \cite{rabaey2003digital,zhang2024dvfo} is a widely-used and classic technique applicable across various hardware platforms, including central processing units (CPUs) and GPUs. Adjusting the operating frequency significantly impacts both processing speed and energy consumption. Typically, increasing frequency enhances processing speed but raises energy consumption, while lowering frequency has the opposite effect. Recently, several studies have explored CPU-GPU joint DVFS \cite{wu2023moc,karzhaubayeva2023cnn} as well as the combined optimization of batching and DVFS \cite{nabavinejad2022coordinated}.

To the best of our knowledge, no prior work has integrated the critical techniques of offloading, batching, and device-edge joint DVFS together in edge-device co-inference. 
In this study, we introduce a novel algorithm called the Joint DVFS, Offloading, and Batching strategy (J-DOB), to minimize total energy consumption while adhering to hard latency constraints in a multi-user co-inference scenario. By determining the minimum latency cost for each user and sorting them accordingly, we establish a set of GPU frequency thresholds along with their corresponding batching sets. Through partition point and edge frequency adjustments, device frequency configuration becomes decoupled and computationally efficient. Ultimately, J-DOB achieves near-optimal DVFS and identical offloading under a greedy batching approach with low complexity.
Extensive experiment results demonstrate significant improvements with J-DOB. When combined with a user-grouping strategy, this algorithm offers a comprehensive solution for energy savings.  By incorporating edge DVFS as a crucial optimization dimension—an advancement over previous work \cite{shi2022multiuser}—we highlight its substantial impact on energy efficiency. Moreover, the proposed algorithm enhances performance even in scenarios that do not utilize edge DVFS.

\section{System Model and Problem Formulation}

\begin{figure}[!t] 
\setlength{\abovecaptionskip}{2pt}
\setlength{\belowcaptionskip}{2pt}
\centering 
\includegraphics[width=0.60\linewidth]{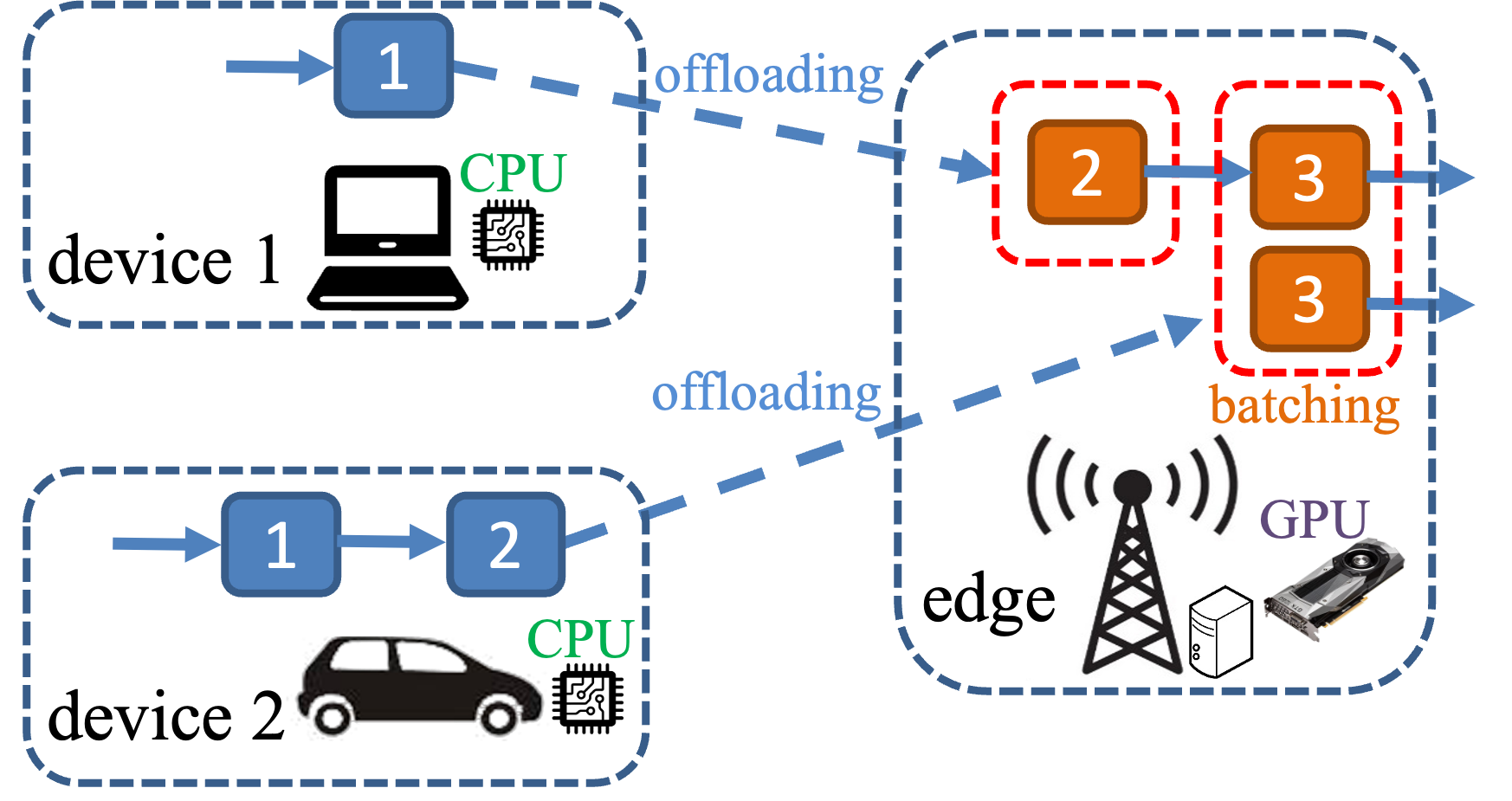} 
\caption{A depiction of edge-device co-inference: The DNN inference task consists of a sequence of sub-tasks. Device 1 offloads sub-task 2 and 3 to the edge, while device 2 offloads sub-task 3. The edge server  processes sub-task 2 with a batch size of one and sub-task 3 with a batch size of two.} 
\label{system}
\vspace{-10pt}
\end{figure}

Consider an edge-device co-inference system with $M$ mobile users equipped with CPU and an edge server that has a single GPU. 
In the following text, we will use the term ``device" and ``user" interchangeably.
As shown in Fig. \ref{system}, all users work with the same pre-trained DNN model to perform inference on their own data. 
Let $T_m^{\mathrm{(d)}}$ denote the latency constraint, or deadline, of user $m$.
The set of mobile users is denoted as $\mathcal{M}=\{1,2,\ldots,M\}$.
Mobile devices are assumed to be capable of performing the full DNN inference locally to meet latency requirements, though this tends to be energy-intensive. 
To reduce energy consumption, users can partition their inference task into smaller sub-tasks (or blocks) and offload certain parts to the edge server.
The edge server, equipped with a GPU, can then batch process these offloaded sub-tasks across multiple users to enhance processing efficiency and throughput\cite{hanhirova2018latency, ren2020accelerating}. 


\subsection{DNN Inference Task Model}
In this paper, the DNN inference task is modeled as \emph{a sequence of sub-tasks}, where each sub-task corresponds to the forward pass of a specific block within the DNN model.
It is assumed that the DNN inference task consists of $N$ sequential sub-tasks. For each sub-task $n \in \{1, \dots, N\}$, let $A_n$ denote its computational workload. The output data size of sub-task $n$, denoted as $O_n$, also serves as the input size for sub-task $n+1$.
Additionally, we introduce $n=0$ as a ``virtual" layer representing the DNN input, where $O_0$ denotes the size of the initial input data and $A_0 = 0$ indicates no computational workload.

\begin{figure}[!t] 
\setlength{\abovecaptionskip}{2pt}
\setlength{\belowcaptionskip}{2pt}
\centering 
\includegraphics[width=0.98\linewidth]{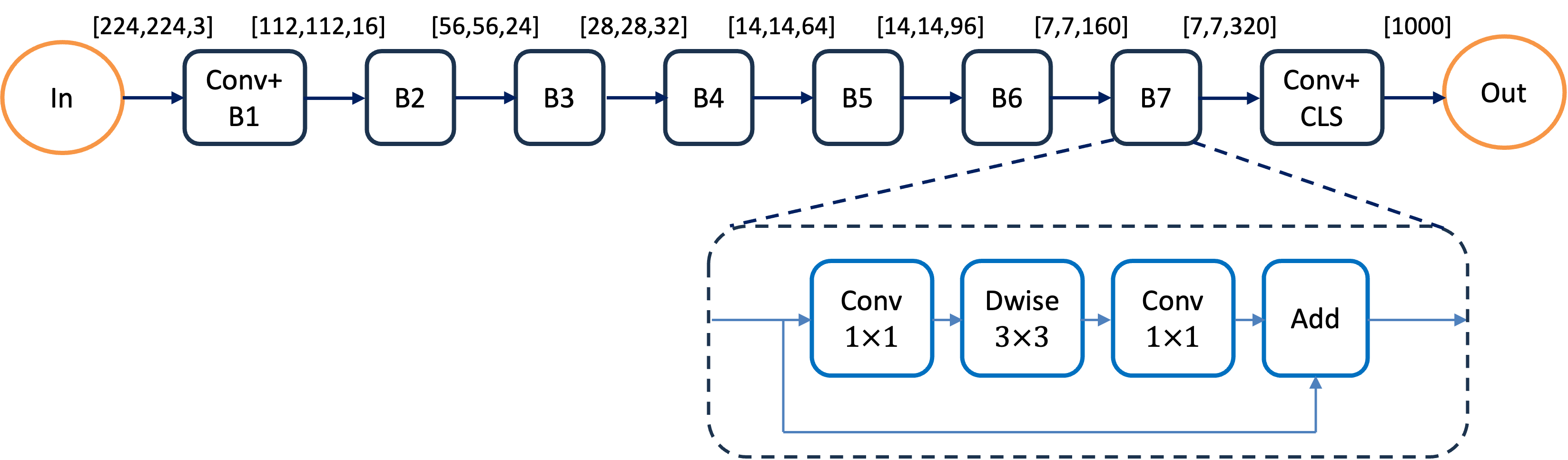} 
\caption{The architecture and partition points of MobileNetV2 used in experiments. Conv, B, and CLS are the abbreviations for convolution module, bottleneck module, and classification module. The architecture of the seventh bottleneck module of MobileNetV2 is illustrated, while the details of other modules can be found in the original paper \cite{sandler2018mobilenetv2}.
The shape of the output data of each sub-task is also demonstrated.} 
\label{arch}
\vspace{-8pt}
\end{figure}

In this work, we focus primarily on MobileNetV2\cite{sandler2018mobilenetv2}, a lightweight DNN model designed for image classification. 
The partitioning strategy for MobileNetV2 employed in this work is illustrated in Fig. \ref{arch}, with partition points set after each block.
Note that finer-grained partitioning at the layer level is also possible; however, this approach would introduce significant complexity, which is beyond the scope of this paper.

\subsection{Co-Inference Model}
\subsubsection{Local Computation}
Let $f_m$ denote the CPU frequency of user $m$, where $f_{m,\text{min}} \leq f_m \leq f_{m,\text{max}}$, and it can be adjusted using the DVFS technique\cite{rabaey2003digital}. 
Given that the computational workload for the $n$-th sub-task is represented by $A_n$, the local computing latency is:
\begin{equation}
    l_{m,n}^\text{cp}(f_m) = \frac{\zeta_m g_n A_n}{f_m}, \label{C1}
\end{equation}
where $\zeta_m$ represents the ratio of CPU cycles to the computational workload (typically measured in FLOPs), and $g_n$ is a factor accounting for block-specific differences.
The corresponding energy consumption for computing is:
\begin{equation}
    e_{m,n}^\text{cp}(f_m) = \kappa_m q_n A_n f_m^2, \label{C2}
\end{equation}
where $\kappa_m$ represents the effective switched capacitance, depending on the chip architecture of device $m$ \cite{yang2020energy}, and $q_n$ reflects differences between blocks.
The mobile devices must be capable of completing the inference task locally within the latency constraint; hence, we have: $\frac{\sum_{n=1}^{N} \zeta_m g_n A_n}{f_{m,\text{max}}} \leq T_m^{\mathrm{(d)}} $.

\subsubsection{Intermediate Data Transmission}
Let $R_m$ denote the transmission rate of user $m$. The transmission latency for uploading the $n$-th sub-task's output data is given by:
\begin{equation}
    l_{m,n}^\text{u} = \frac{O_n}{R_m}. \label{C3}
\end{equation}
The corresponding transmission energy consumption is:
\begin{equation}
    e_{m,n}^\text{u} = l_{m,n}^\text{u} p_m^\text{u}, \label{C4}
\end{equation}
where $p_m^\text{u}$ represents the power consumption of the transmitter. 
Since the DNN output data size is relatively small, the download time is minimal and thus omitted in this analysis.

\subsection{Batch Processing Model}
\begin{figure}[!t]
\setlength{\abovecaptionskip}{1pt}
\setlength{\belowcaptionskip}{1pt}
\centering
\subfloat[\footnotesize{latency}]{
\includegraphics[width=0.54\linewidth]{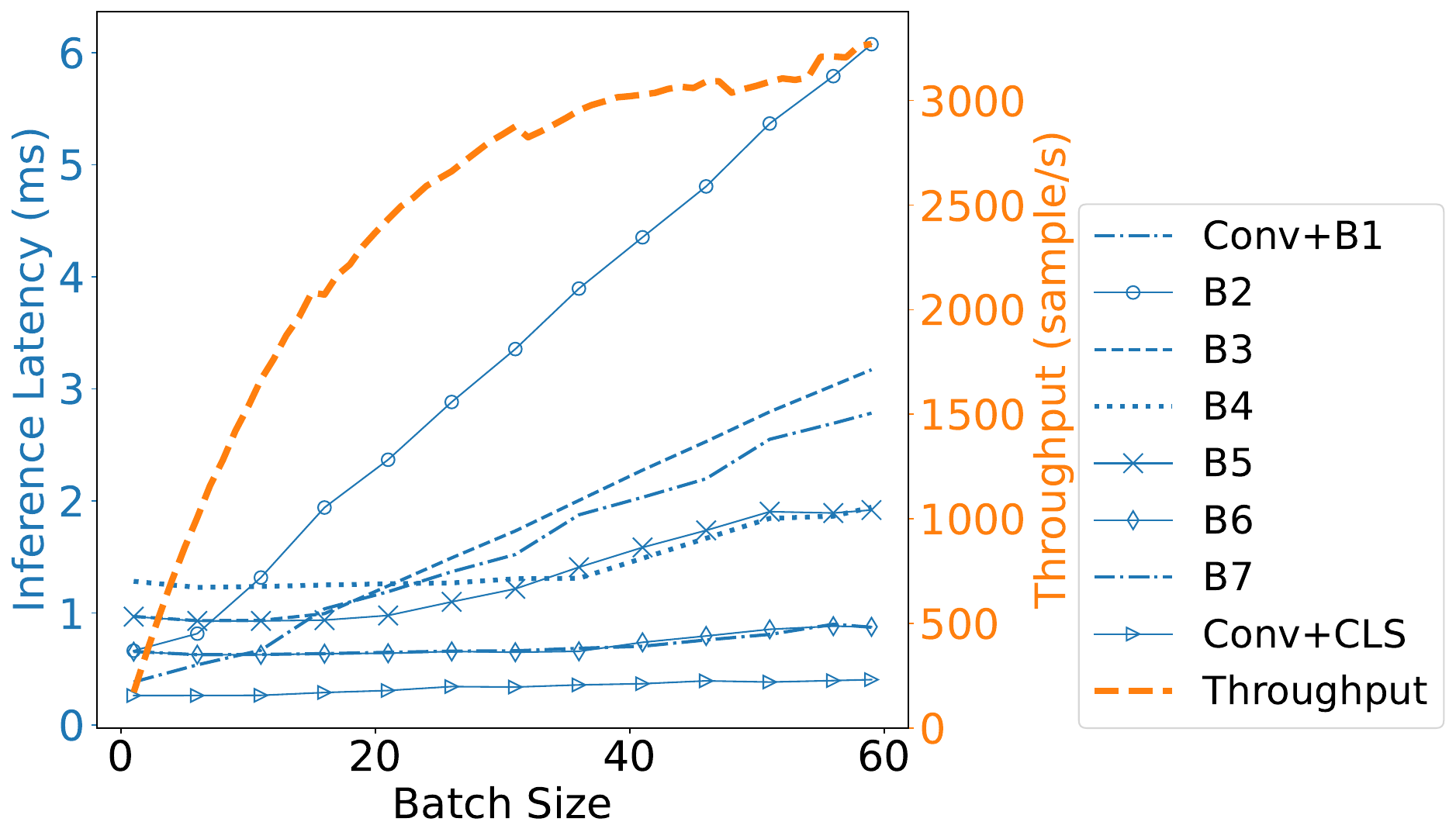}
\label{xx_a}}
\hspace{-9pt}
\subfloat[\footnotesize{energy consumption}]{
\includegraphics[width=0.42\linewidth]{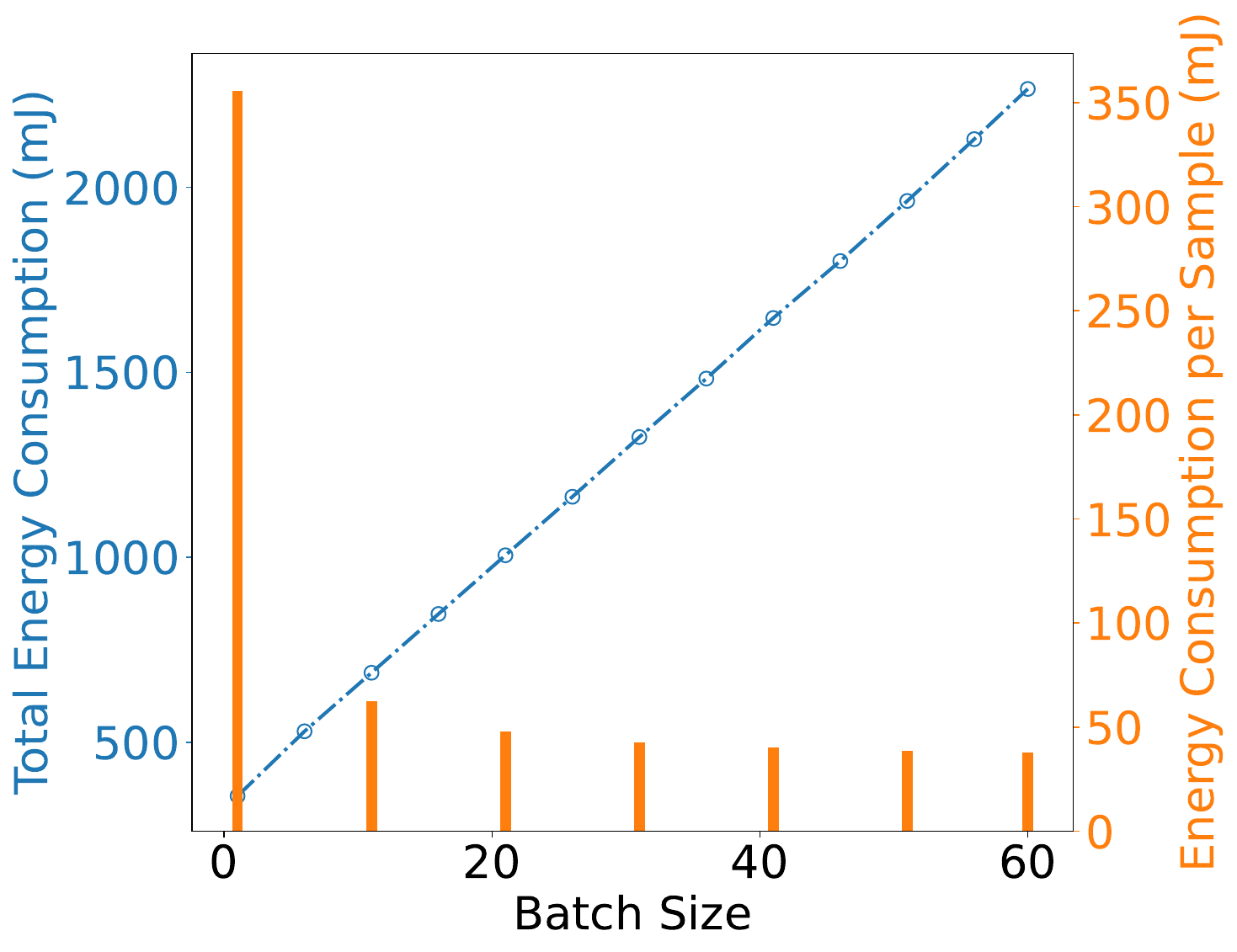}
\label{xx_b}}
\vspace{2pt}
\caption{Profiling results for  inference latency and energy consumption of MobileNetV2 w.r.t. batch size on an NVIDIA RTX3090\cite{3090}.}
\label{xx} 
\end{figure}

We define a set of functions $L_n(\cdot)$ and $E_n(\cdot)$ to represent the computing latency and energy consumption of the edge server for each sub-task, in terms of the batch size $b$ and the GPU core frequency $f_\mathrm{e}$:

\begin{equation}
\begin{aligned}
    &~ L_n(f_\mathrm{e},b) = d_n(b) A_n \frac{1}{f_\mathrm{e}},\\
    &~ E_n(f_\mathrm{e},b) = c_n(b) A_n f_\mathrm{e}^2,
    \label{C12}
\end{aligned}
\end{equation}
where $d_n(\cdot)$ and $c_n(\cdot)$ are functions that capture block-specific variations in relation to the batch size. The GPU frequency $f_\mathrm{e}$ can be adjusted using DVFS, within the range $f_{\mathrm{e}, \min } \leq f_\mathrm{e} \leq f_{\mathrm{e}, \max }$ .
The profiling results for MobileNetV2 on a GPU server, as shown in Fig.~\ref{xx}\subref{xx_a} and Fig.~\ref{xx}\subref{xx_b}, indicate that both inference latency and energy consumption increase with the batch size. Meanwhile, latency and energy per sample decrease as the batch size grows.


\subsection{Problem Statement and Solution Overview}
In this work, we aim to minimize the total energy consumption under hard inference latency constraints.
The time constraints in this problem include synchronization constraints (ensuring the upload of all necessary intermediate data before batch processing any sub-task), sequence constraints (sub-tasks must be processed in order), and hard deadline constraints.
This problem extends the framework in \cite{shi2022multiuser} by incorporating GPU frequency scaling as an additional design variable, incorporating the GPU available time in the constraint, and accounting for edge energy in the objective function. Due to space limitations, we do not present the full formulation of the initial optimization problem.

The interdependencies among partitioning and offloading decisions, frequency control, and batch scheduling (when and how to batch) make the problem highly complex. As described in \cite{shi2022multiuser}, this problem is classified as a mixed-integer nonlinear programming problem (MINLP), which is NP-hard \cite{wolsey2007mixed}, and it expands rapidly in solution space as the problem scales.
Following a similar approach to \cite{shi2022multiuser}, our problem is decomposed into two parts: an outer module that groups users by deadline similarity, and an inner module that jointly optimizes partition points (offloading strategy), edge and device frequencies, and the batch scheduling scheme. In this work, we mainly focus on the inner module. The problem formulation and algorithm design are detailed in the following section.



\section{Joint DVFS, Offloading and Batching within a Given Group}
The inner optimization problem is still complicated and includes a vast solution space.
Therefore, we simplify the solution space by adding two more constraints: (1) All offloading tasks share a single common partition point, which is called \emph{identical offloading}. (2) The edge server carries out greedy batching, which means that it will bundle all available identical sub-tasks into a whole batch.

Given a set of devices, $\mathcal{M}' \subset \mathcal{M}$, we can decompose it into an offloading set $\mathcal{M'_{\mathrm{o}}}$ and a local computing set $\mathcal{M'_{\mathrm{l}}}$.
We introduce $t_{\mathrm{free}}$, representing the time at which the GPU becomes available.
What we need to do is to design an optimal identical offloading and DVFS scheme under greedy batching.
Specifically, for the offloading part, we need to determine the offloading set $\mathcal{M'_{\mathrm{o}}}$ as well as the identical partition point $\tilde{n} \in \{0,1,\ldots,N\}$, where $\tilde{n}=n \; (0<n<N)$ represents  offloading right after the $n$-th block, $\tilde{n}=0$ represents whole task offloading, and $\tilde{n}=N$ means local computing.

The simplified inner optimization problem is as follows:
\begin{align}
    \min_{ \mathcal{M'_{\mathrm{o}}},\tilde{n},f_m,f_\mathrm{e}} &~ \sum_{m \in \mathcal{M'_{\mathrm{o}}}} \sum_{n=1}^{\tilde{n}} \kappa_m q_n A_n f_m^2 + \sum_{m \in \mathcal{M'_{\mathrm{o}}}} \frac{O_{\tilde{n}}}{R_m} p_m^{\mathrm{u}} \nonumber \\
    &+\sum_{n=\tilde{n}+1}^N c_n(B_{\mathrm{o}})A_n{f^2_\mathrm{e}} 
    + \sum_{m \in \mathcal{M'_{\mathrm{l}}}} \sum_{n=1}^{N} \kappa_m q_n A_n f_m^2 \tag{P1}\label{P1}\\
\mathrm { s.t. } &~ t_{\mathrm{free}}+\sum_{\tilde{n}+1}^{N}d_n(B_{\mathrm{o}})A_n\frac{1}{f_\mathrm{e}} \leq l_{\mathrm{o}},\label{D6}\\
&~ \frac{\sum_{n=1}^{\tilde{n}} \zeta_m g_n A_n}{f_m}+\frac{O_{\tilde{n}}}{R_m}+\sum_{\tilde{n}+1}^{N}d_n(B_{\mathrm{o}})A_n\frac{1}{f_\mathrm{e}} \leq l_{\mathrm{o}}, \nonumber \\
&\quad \quad \quad \quad \quad \quad \quad \quad \quad \quad \quad \forall m \in \mathcal{M'_{\mathrm{o}}},\label{D7}\\
&~ \frac{\sum_{n=1}^{N} \zeta_m g_n A_n}{f_m} \leq T_m^{\mathrm{(d)}}, \quad \forall m \in \mathcal{M'_{\mathrm{l}}},\label{D8}\\
&~ \tilde{n} \in \{0,1,\ldots,N\},\label{D9}\\
&~ l_{\mathrm{o}} = \min_{m \in \mathcal{M'_{\mathrm{o}}}} \{T_m^{\mathrm{(d)}}\},\label{D10}\\
&~ \mathcal{M'_{\mathrm{o}}} \subset \mathcal{M}', \label{D10plus}\\
&~ B_{\mathrm{o}} = |\mathcal{M'_{\mathrm{o}}}|,\label{D11}\\
&~ \mathcal{M'_{\mathrm{l}}} = \mathcal{M}' \setminus \mathcal{M'_{\mathrm{o}}},\label{D12}\\
&~ f_{m, \min } \leq f_m \leq f_{m, \max }, \quad \forall m \in \mathcal{M}',\label{D13}\\
&~ f_{\mathrm{e}, \min } \leq f_\mathrm{e} \leq f_{\mathrm{e}, \max }. \label{D14}
\end{align}

Eq.~(\ref{D6}) is the GPU occupation constraint at the edge. 
Eq.~(\ref{D7}) is the latency constraint for device-edge co-inference. 
The deadline for batching, $l_{\mathrm{o}}$ (specified in Eq.~(\ref{D10})), is the tightest deadline in the offloading set.
Eq.~(\ref{D8}) is the inference latency constraint for local (device-only) computing.
Due to greedy batching, the batch size $B_{\mathrm{o}}$ is exactly the size of the offloading set, as shown in Eq.~(\ref{D11}).

Note that $\mathcal{M'_{\mathrm{o}}}$ has $2^{|\mathcal{M'}|}$ possible choices and $\tilde{n}$ has $N+1$ choices.
Consequently, obtaining the optimal solution within polynomial complexity appears infeasible.
However, it should be also noticed that the feasible solution space is much smaller than the total solution space.
Based on this point, we propose Algorithm~\ref{alg1}, termed the Joint DVFS, Offloading, and Batching strategy (J-DOB) to solve this problem effectively.
Several notations are made for simplicity:
$u_{\tilde{n}} \triangleq \sum_{n=0}^{\tilde{n}} q_n A_n, \; \psi_{\tilde{n}}(B_{\mathrm{o}}) \triangleq \sum_{n=\tilde{n}+1}^{N} c_n(B_{\mathrm{o}}) A_n, \; v_{\tilde{n}} \triangleq \sum_{n=0}^{\tilde{n}} g_n A_n$, and  $\phi_{\tilde{n}}(B_{\mathrm{o}}) \triangleq \sum_{n=\tilde{n}+1}^{N} d_n(B_{\mathrm{o}}) A_n
$.

To start with, since the number of partition points is usually limited, we exhaustively search through all possibilities in the outer loop to find the optimal solution (see Alg.~\ref{alg1}, line 3). Within each iteration, we jointly design the batching and DVFS schemes. The core challenge of this problem lies in maintaining feasibility due to strict latency constraints and frequency range limits. Notably, Eq. (\ref{D7}) serves as the primary constraint linking edge frequency, device frequencies, and batching. By rearranging Eq. (\ref{D7}), we derive the required edge GPU frequency: 
\begin{equation}
    f_\mathrm{e} \geq \frac{\phi_{\tilde{n}}(B_{\mathrm{o}})}{l_{\mathrm{o}} - \frac{O_{\tilde{n}}}{R_m} - \frac{\zeta_m v_{\tilde{n}}}{f_m}} \geq 0, \quad \forall m \in \mathcal{M'_{\mathrm{o}}}.
\end{equation}

We then seek to determine the minimum edge frequency that enables user $m$ to offload its sub-tasks after the $\tilde{n}$-th block. This is challenging because both $B_{\mathrm{o}}$ and $l_{\mathrm{o}}$ are determined by the specific set $\mathcal{M'_{\mathrm{o}}}$, creating inter-dependencies in the computation among users. However, we identify a variable that effectively addresses this issue. Define $\gamma_m^{(\tilde{n})}$ as the minimum latency cost for user $m$ given the partition point $\tilde{n}$:
\begin{equation}
    \gamma_m^{(\tilde{n})} \triangleq \frac{O_{\tilde{n}}}{R_m}+\frac{ \zeta_m v_{\tilde{n}}}{f_{m,\max}}, \quad \forall m \in \mathcal{M'}.
\end{equation}
A higher value of $\gamma_m^{(\tilde{n})}$ implies a smaller latency budget for batching and therefore a higher required edge frequency. Based on this, we sort $\mathcal{M}'$ into $\mathrm{list}(\mathcal{M}')$ in descending order of $\gamma_m^{(\tilde{n})}$ (as shown in Alg.~\ref{alg1}, line 5). The corresponding minimum edge frequency, which serves as a threshold, can then be calculated.


For the $i$-th user in $\mathrm{list}(\mathcal{M}')$, it leads to a threshold frequency of 
\begin{equation}
    f_\mathrm{e}^{\mathrm{th},i} = \frac{\phi_{\tilde{n}}(B-i+1)}{\min_{m \in \mathrm{list}(\mathcal{M}')[i:B]} \{T_m^{\mathrm{(d)}}\}-\gamma_m^{(\tilde{n})}}.\label{fth}
\end{equation}


\begin{algorithm}
\SetAlgoLined
\caption{Joint DVFS, Offloading, and Batching Strategy within a Given Group (J-DOB)}
\label{alg1}

\KwIn {$\mathcal{M'}, \; t_{\mathrm{free}}, \; \rho$.}
\KwRequire {$\min \limits_{m \in \mathcal{M'}} \{T_m^{\mathrm{(d)}}\} \geq t_{\mathrm{free}}$.}

\KwOut{$(E_*, \; t_{\mathrm{free},*}, \; \mathcal{X}_*)$.}
{$B \leftarrow |\mathcal{M'}|$.}

{Initialize $E_* \leftarrow +\infty, \; t_{\mathrm{free},*} \leftarrow t_{\mathrm{free}}, \; \mathcal{X}_* \leftarrow \emptyset$.}

\For (\tcp*[f]{traverse partition points}){$\tilde{n} \leftarrow 0$ to $N$}
{

{Compute $\gamma_m^{(\tilde{n})} \leftarrow \frac{O_{\tilde{n}}}{R_m}+\frac{ \zeta_m v_{\tilde{n}}}{f_{m,\max}}, \forall m \in \mathcal{M'}$.}

{Sort $\mathcal{M}'$ to $\mathrm{list}(\mathcal{M}')$ by the descending order of $\gamma_m^{(\tilde{n})}$, and let $\mathrm{list}(\mathcal{M}')[i]$ be the $i$-th item of $\mathrm{list}(\mathcal{M}')$.}

{Compute the edge frequency thresholds $f_\mathrm{e}^{\mathrm{th},i}, \forall i \in \{1,2,\ldots,B\}$ according to Eq. (\ref{fth}).}
{Call Alg.~\ref{alg2} to obtain the strategy $\mathcal{X}_*^{(\tilde{n})}$, which results in the energy consumption of $E_*^{(\tilde{n})}$ and occupies the edge server until $t_{\mathrm{free},*}^{(\tilde{n})}$, under the given identical offloading point $\tilde{n}$.}

\If{$E_*^{(\tilde{n})}<E_*$}{

{$E_* \leftarrow E_*^{(\tilde{n})}, \; t_{\mathrm{free},*} \leftarrow t_{\mathrm{free},*}^{(\tilde{n})}, \; \mathcal{X}_* \leftarrow \mathcal{X}_*^{(\tilde{n})}$.}
}
}

\end{algorithm}

It is noted that the offloading set (also the greedy batching set) is updated as well. 
If $f_\mathrm{e} \geq f_\mathrm{e}^{\mathrm{th},i}$, then user $i$ should be in the offloading set $\mathcal{M'_{\mathrm{o}}}$ according to the greedy batching rule.
Otherwise, it should be removed.
It is easy to validate that $\{f_\mathrm{e}^{\mathrm{th},i}\}_{i \in \mathrm{list}(\mathcal{M}')}$ is a non-increasing sequence.
Therefore, in Alg.~\ref{alg2} (which is invoked by Alg.~\ref{alg1}), the edge frequency is swept from $f_{\mathrm{e}, \max }$ to $f_{\mathrm{e}, \min }$ with a step size of $\rho$, and the offloading set $\mathcal{M'_{\mathrm{o}}}$ updates accordingly with linear complexity (see Alg.~\ref{alg2}, line 7-12).

After fixing $\tilde{n}$, $f_\mathrm{e}$, and $\mathcal{M'_{\mathrm{o}}}$, problem (\ref{P1}) simplifies into a convex optimization problem that can be decoupled for each $m \in \mathcal{M'}$.
Define $\Gamma_m(\tilde{n},\mathcal{M'_{\mathrm{o}}},f_\mathrm{e})$ as 
\begin{align}
\Gamma_m(\tilde{n},\mathcal{M'_{\mathrm{o}}},f_\mathrm{e})=
\begin{cases}
    \frac{\zeta_m v_{\tilde{n}}}{l_{\mathrm{o}}-\frac{O_{\tilde{n}}}{R_m}-\frac{\phi_{\tilde{n}}(B_{\mathrm{o}})}{f_\mathrm{e}}}, &m \in \mathcal{M'_{\mathrm{o}}}\\
    \frac{\zeta_m v_{N}}{T_m^{\mathrm{(d)}}}, &m \in \mathcal{M'_{\mathrm{l}}}
\end{cases}.
\end{align}
The optimal device frequency $f_{m}^*$ can be easily obtained, along with the corresponding energy consumption $E^*$ and the end of GPU occupation $t_{\mathrm{free}}^*$:
\begin{align}
    f_{m}^* =& \min\big\{\max \big\{\Gamma_m(\tilde{n},\mathcal{M'_{\mathrm{o}}}, f_\mathrm{e}),f_{m,\min} \big\}, f_{m,\max} \big\}, \label{D20}\\
    E^* =& \sum_{m \in \mathcal{M'_{\mathrm{o}}}} \left(\kappa_m u_{\tilde{n}} (f_m^*)^2 + \frac{O_{\tilde{n}}}{R_m} p_m^{\mathrm{u}} \right) \nonumber \\
    +& \sum_{m \in \mathcal{M'_{\mathrm{l}}}} \kappa_m u_N (f_m^*)^2
    +\psi_{\tilde{n}}(B_{\mathrm{o}}){f^2_\mathrm{e}}, \label{D21}\\
    t_{\mathrm{free}}^* =& \max \big\{t_{\mathrm{free}}, \max_{m \in \mathcal{M'_{\mathrm{o}}}} \big\{ \frac{ \zeta_m v_{\tilde{n}}}{f_m^*}+\frac{O_{\tilde{n}}}{R_m} \big\} \big\} + \frac{\phi_{\tilde{n}}(B_{\mathrm{o}})}{f_\mathrm{e}}. \label{D22}
\end{align}

Suppose the number of swept edge frequency points is $k$. The time complexity of the proposed J-DOB algorithm is then $\mathcal{O}(kNM\log{M})$, where the complexity in $M$ is primarily due to the sorting procedure (see Alg. \ref{alg1}, line 5). 
This algorithm effectively addresses the potentially high complexity of handling the combinatorial problem regarding $\mathcal{M'}$.

\begin{algorithm}
\SetAlgoLined
\caption{Joint Edge and Device DVFS Optimization under Identical Offloading and Greedy Batching} 
\label{alg2}
\KwIn {$\tilde{n}, \; \mathrm{list}(\mathcal{M}'), \; \{f_\mathrm{e}^{\mathrm{th},i}\}_{i \in \{1,2,\ldots,B\}}, \; t_{\mathrm{free}}, \; \rho, \; B$.}
\KwOut{$(E_*^{(\tilde{n})}, \; t_{\mathrm{free},*}^{(\tilde{n})}, \; \mathcal{X}_*^{(\tilde{n})})$.}
{Initialize $E_*^{(\tilde{n})} \leftarrow +\infty, \; t_{\mathrm{free},*}^{(\tilde{n})} \leftarrow t_{\mathrm{free}}, \; \mathcal{X}_*^{(\tilde{n})} \leftarrow \emptyset$.}

{Set $\hat{i} \leftarrow \min\{i \vert i \in \{1,2,...B\}, f_\mathrm{e}^{\mathrm{th},i} \geq 0\}$.}

{Initialize $\mathcal{M'_{\mathrm{o}}} \leftarrow \mathrm{list}(\mathcal{M}')[\hat{i}:B]$  if $\hat{i} \neq \text{NAN}$, and $\mathcal{M'_{\mathrm{o}}} \leftarrow \emptyset$ otherwise.} 

{$\mathcal{M'_{\mathrm{l}}} \leftarrow \mathcal{M'} \setminus \mathcal{M'_{\mathrm{o}}}, \; l_{\mathrm{o}} \leftarrow \min \limits_{m \in \mathcal{M'_{\mathrm{o}}}} \{T_m^{\mathrm{(d)}}\}, \;
B_{\mathrm{o}} \leftarrow |\mathcal{M'_{\mathrm{o}}}|$.}

{Initialize $f_{\mathrm{e}} \leftarrow f_{\mathrm{e},\max}$.}

\While (\tcp*[f]{sweep edge frequency}){$f_\mathrm{e} \geq f_{\mathrm{e}, \min}$}{
\If (\tcp*[f]{update greedy batching set}){$\hat{i} \neq \mathrm{NaN}$}{
\While {$\hat{i} \leq B$ and $f_\mathrm{e} < f_\mathrm{e}^{\mathrm{th},\hat{i}}$}{
{Update $\mathcal{M'_{\mathrm{o}}} \leftarrow \mathcal{M'_{\mathrm{o}}} \setminus \{\mathrm{list}(\mathcal{M}')[\hat{i}]\},$}
{$ \mathcal{M'_{\mathrm{l}}} \leftarrow \mathcal{M'_{\mathrm{l}}} \cup \{\mathrm{list}(\mathcal{M}')[\hat{i}]\},$} 
{$l_{\mathrm{o}} \leftarrow \min \limits_{m \in \mathcal{M'_{\mathrm{o}}}} \{T_m^{\mathrm{(d)}}\}, \; 
B_{\mathrm{o}} \leftarrow |\mathcal{M'_{\mathrm{o}}}|$.}

{Update $\hat{i} \leftarrow \hat{i} + 1$.}
}
}
\If (\tcp*[f]{optimal device dvfs}){$f_\mathrm{e} \geq \frac{\phi_{\tilde{n}}(B_{\mathrm{o}})}{l_{\mathrm{o}}-t_{\mathrm{free}}}$}
{
{$(\{f_{m}^*\}_{m \in \mathcal{M}'}, E^*, t^*_{\mathrm{free}}) \leftarrow \text{Eq. \eqref{D20}-\eqref{D22}}$.}

{${\tilde{n}}_m \leftarrow \tilde{n}, \forall m \in \mathcal{M'_{\mathrm{o}}}$.}

\If {$E^* < E_*^{(\tilde{n})}$}
{$E_*^{(\tilde{n})} \leftarrow E^*, \; t_{\mathrm{free},*}^{(\tilde{n})} \leftarrow t^*_{\mathrm{free}}, \;  \mathcal{X}_*^{(\tilde{n})} \leftarrow (\mathcal{M'_{\mathrm{o}}}, \mathcal{M'_{\mathrm{l}}}, \{{\tilde{n}}_m\}_{m \in \mathcal{M'_{\mathrm{o}}}}, \{f_{m}^*\}_{m \in \mathcal{M'}}, f_{\mathrm{e}})$.}
}
\eIf {$\mathcal{M'_{\mathrm{o}}} = \emptyset$}{{\Break}}{{$f_\mathrm{e} \leftarrow f_\mathrm{e} - \rho$.}}
}
\end{algorithm}

\section{Experiment Results}
We evaluate the proposed algorithm on MobileNetV2\cite{sandler2018mobilenetv2}, whose architecture and sub-task partitioning are shown in Fig. \ref{arch}.
The profiling results with different batch sizes on an NVIDIA RTX3090\cite{3090} are shown in Fig.~\ref{xx}\subref{xx_a} and Fig.~\ref{xx}\subref{xx_b}.
Moreover, the computing load of each sub-task is  profiled using torchsummaryX.
The system parameters are listed in Table \ref{parameter_offline}.
The transmission rate is $R_m=W_m \text{log}_2(1+ \text{SNR})$.
We define $\alpha_m$ as the ratio of the local inference latency to the edge inference latency with batch size of $1$, both at the maximum frequency.
We define $\eta_m$ as the ratio of the local inference power level to the edge inference power level with batch size of $1$, both at the maximum frequency.
Other parameters can be obtained by fitting the energy and latency statistics. 
Another parameter, $\beta$, represents the ratio between the task deadline and the shortest local inference latency, minus one: $\beta_m \triangleq \frac{T_m^{\mathrm{(d)}}}{\sum_{n=1}^{N} l^\text{cp}_{m,n}(f_{m,\text{max}})} - 1$,
which indicates the tightness of the latency constraint.

\begin{table}[!t]
\setlength{\abovecaptionskip}{2pt}
\setlength{\belowcaptionskip}{2pt}
\caption{System Parameters}
\label{parameter_offline}
\begin{center}
\begin{tabular}{|c |c |c |c|}
\hline
\textbf{Parameter} & \textbf{Value} & \textbf{Parameter} & \textbf{Value}\\
\hline
SNR & 30 dB & $\alpha_m$ & 1   \\
$W_m$ & 10 MHz & $\eta_m$ & 0.6 \\
$g_n$ & 1 & $q_n$ & 1 \\
$p^\text{u}_m$ & 1 W & $\rho$ & 0.03 GHz  \\
$f_{m,\text{min}}$ & 1.5 GHz & $f_{m,\text{max}}$ & 2.6 GHz  \\
$f_{\mathrm{e}, \min }$ & 0.2 GHz & $f_{\mathrm{e}, \max }$ & 2.1 GHz \\
\hline
\end{tabular}
\end{center}
\vspace{-10pt}
\end{table}



The performance of J-DOB is compared with the following benchmarks: (i) Local computing (LC) (ii) Independent partitioning and same sub-task aggregating (IP-SSA)\cite{shi2022multiuser} (iii) J-DOB without edge DVFS (J-DOB w/o edge DVFS) and (iv) J-DOB with binary offloading (J-DOB binary), i.e., J-DOB with $\tilde{n} \in \{0,N\}$ .
In both IP-SSA\cite{shi2022multiuser} and J-DOB w/o edge DVFS, the GPU frequency is fixed at $f_{\mathrm{e}, \max }$, while device DVFS is maintained across all methods.
To thoroughly evaluate the performance of the proposed J-DOB algorithm, we conduct experiments in two scenarios: one with an identical deadline and another with different deadlines. 

\subsection{Identical Deadline}

\begin{figure}[!t]
\setlength{\abovecaptionskip}{1pt}
\setlength{\belowcaptionskip}{1pt}
\centering
\subfloat[\footnotesize{$\beta = 2.13$ ($T_m^{\mathrm{(d)}} = 10$ ms)}]{
\includegraphics[width=0.46\linewidth]{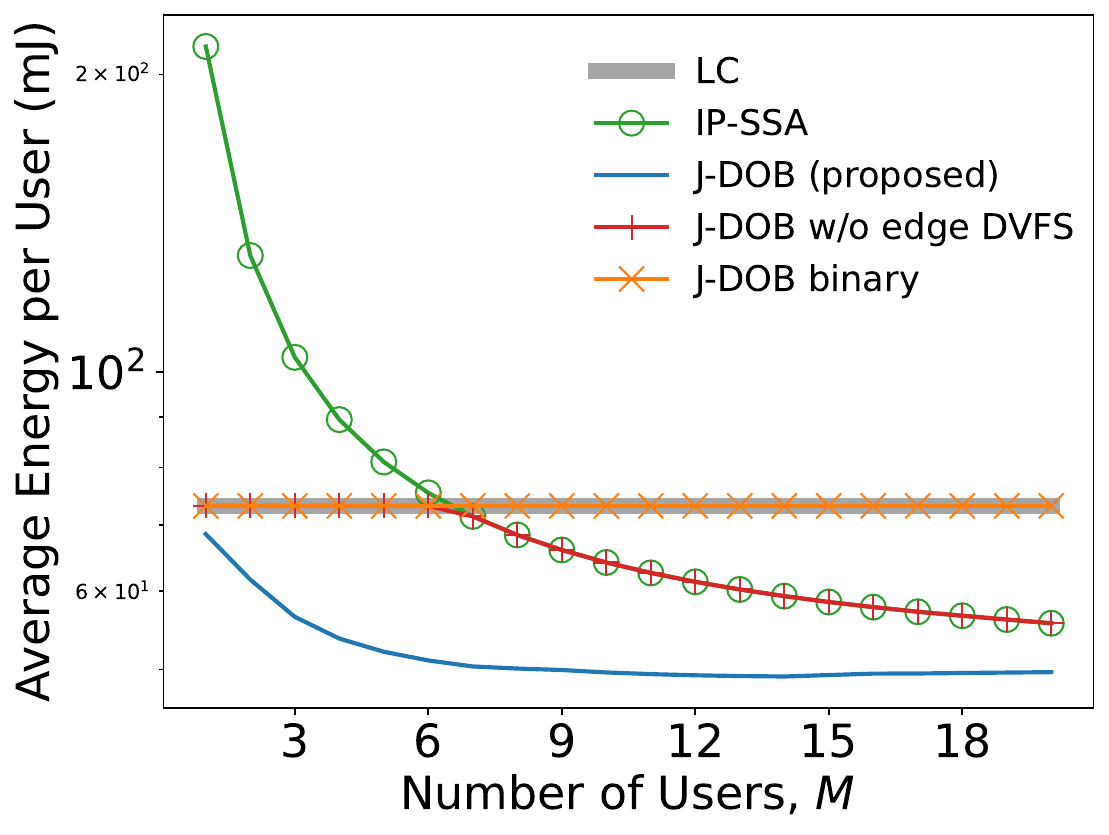}
\label{o_b_a}}
\vspace{0pt}
\subfloat[\footnotesize{$\beta = 30.25$ ($T_m^{\mathrm{(d)}} = 100$ ms)}]{
\includegraphics[width=0.46\linewidth]{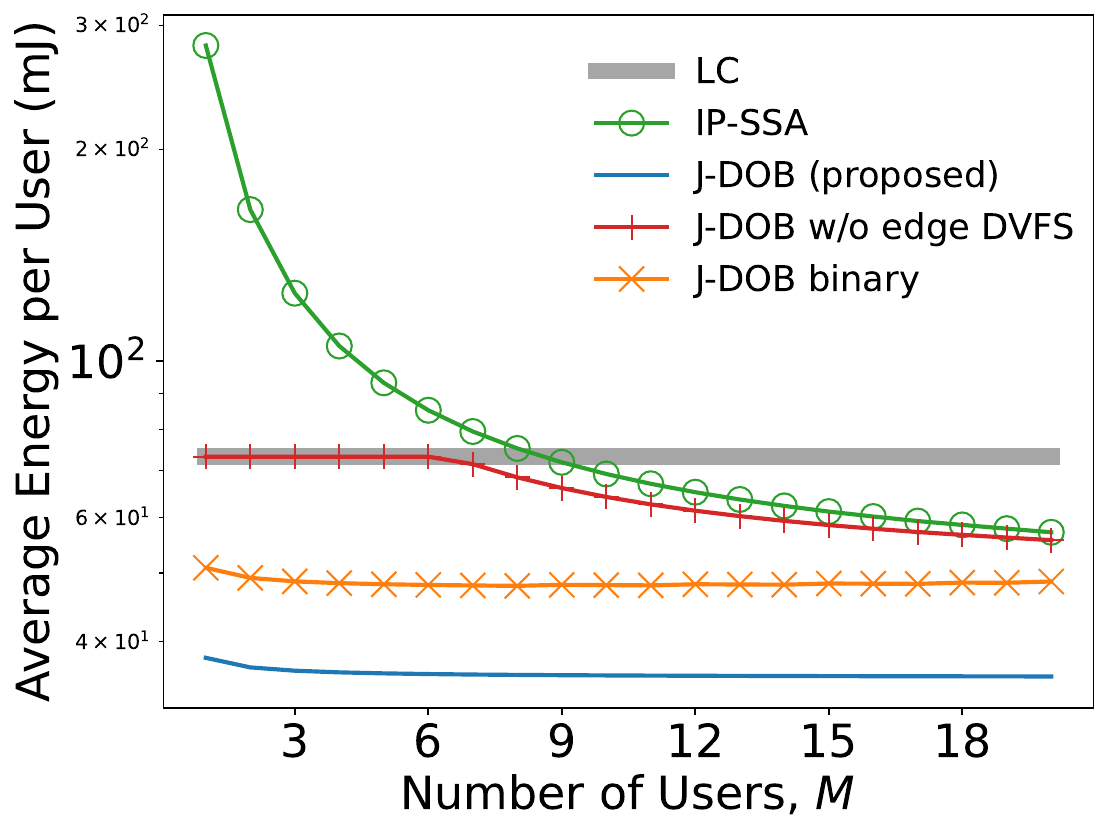}
\label{o_b_b}}
\vspace{2pt}
\caption{Average energy consumption per user v.s. the number of users under identical deadlines. Fig. \ref{o_b} (a) and (b) show the results for different identical deadline values, corresponding to $\beta=2.13$ and $\beta =30.25$, respectively.}
\label{o_b} 
\end{figure}

For the scenario where all users have an identical deadline, we evaluate the proposed J-DOB algorithm under $\beta=2.13$ and $\beta=30.25$, respectively.
Fig. \ref{o_b}(a) and Fig. \ref{o_b}(b) show the average energy consumption per user with respect to the number of users.
Firstly, J-DOB, J-DOB without edge DVFS, and J-DOB binary consistently consume equal or less energy compared to LC.
Notably, J-DOB demonstrates optimal performance in all cases, achieving a maximum energy reduction of 32.8\% and 51.3\% under  $\beta = 2.13$  and  $\beta = 30.25$ , respectively, when compared to LC.
IP-SSA performs poorly with small batch sizes, as GPU energy efficiency is lower than that of CPU in such cases.
When comparing J-DOB without edge DVFS to IP-SSA, the former shows a clear advantage when  $M$  is small and even maintains some superiority as  $M$ increases (see Fig. \ref{o_b}(b)).
This suggests that J-DOB achieves significant improvements even in the original configuration of \cite{shi2022multiuser} without the inclusion of edge DVFS.

\subsection{Different Deadlines}
In experiments of the subsection, the deadlines for users can be different.
We first define the range of deadlines (represented by the range of $\beta$), and then independently set the deadlines based on a uniform distribution.
The dynamic programming algorithm for optimal grouping (OG)\cite{shi2022multiuser} is selected as the outer grouping algorithm.
The inner algorithm, e.g. J-DOB or any other benchmark, is called by OG\cite{shi2022multiuser} to generate a complete strategy with multiple batches.

\begin{figure}[!t]
\setlength{\abovecaptionskip}{1pt}
\setlength{\belowcaptionskip}{1pt}
\centering
\subfloat[\footnotesize{$M=10$}]{
\includegraphics[width=0.46\linewidth]{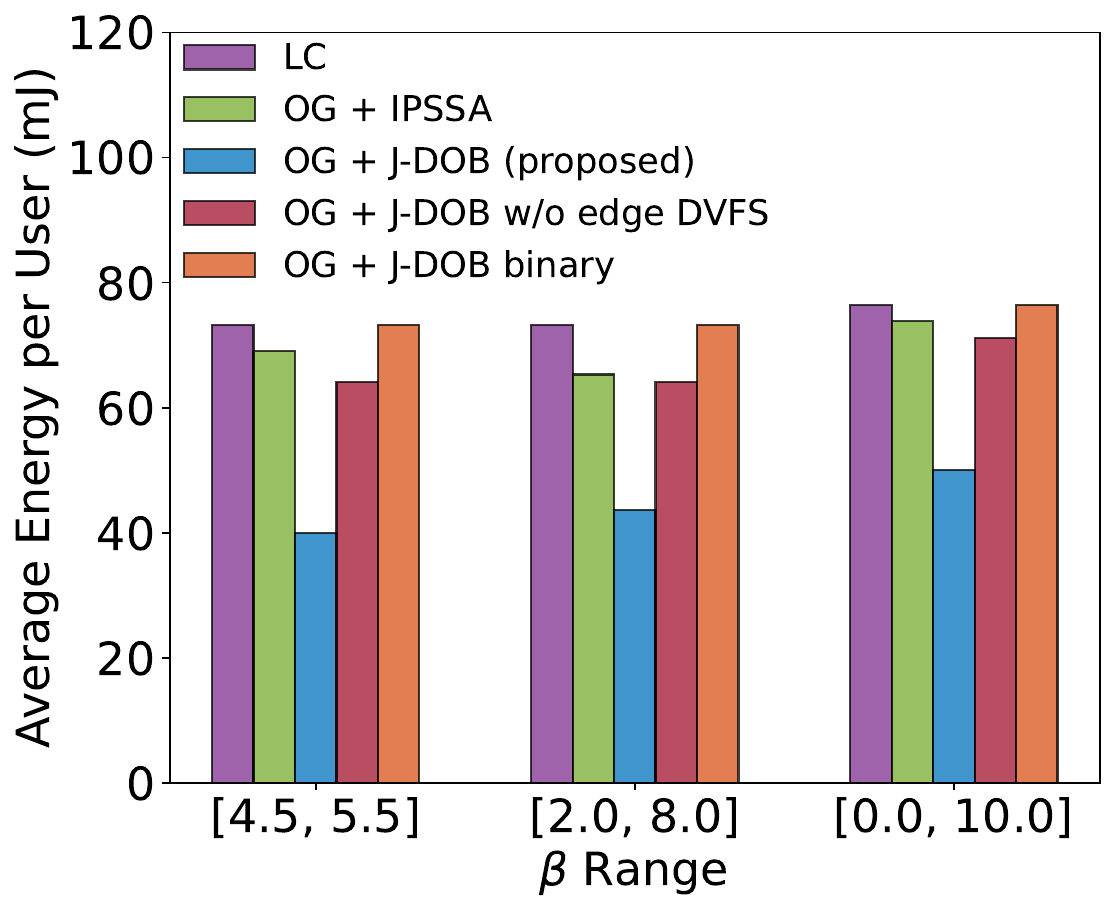}
\label{ee_b_a}}
\vspace{0pt}
\subfloat[\footnotesize{$M=20$}]{
\includegraphics[width=0.46\linewidth]{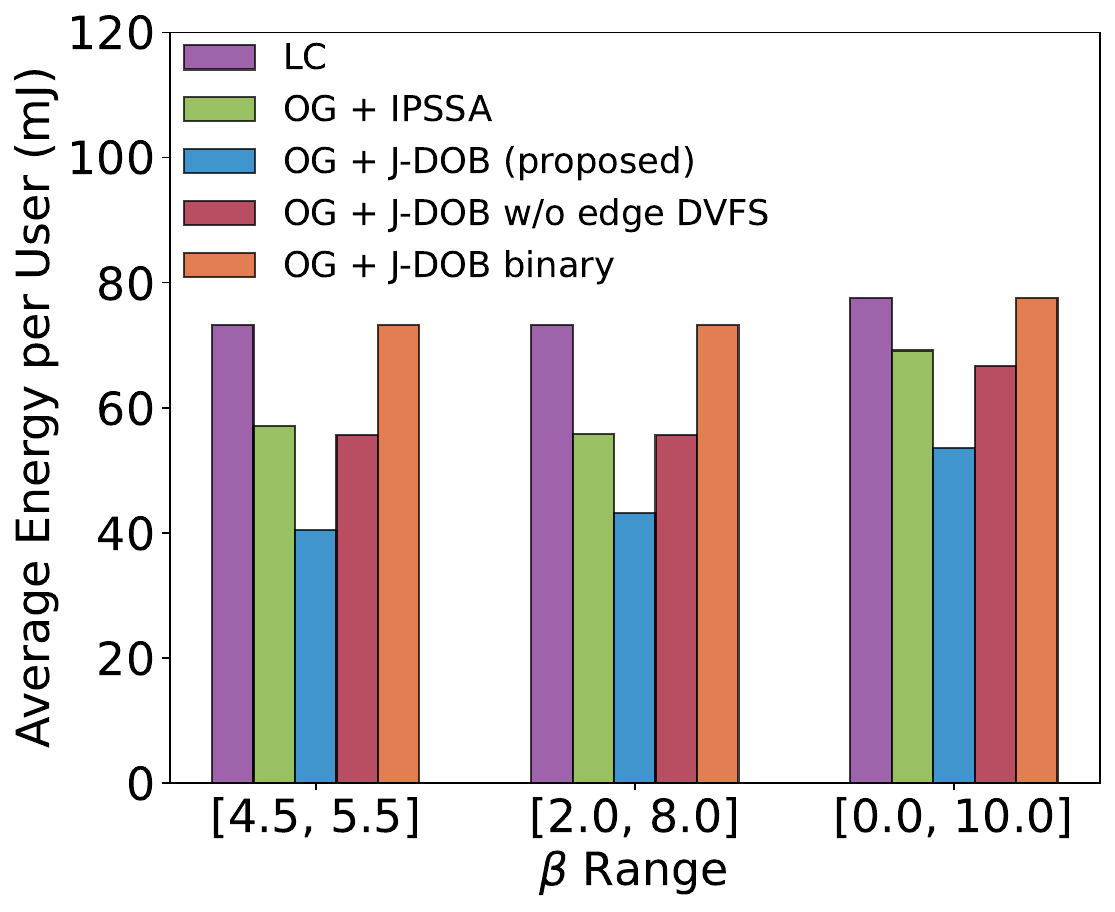}
\label{ee_b_b}}
\vspace{2pt}
\caption{Average energy consumption per user v.s. the range of $\beta$ (corresponding to the range of deadlines). Fig. \ref{ee_b} (a) and (b) show the results under $M=10$ and $M=20$, respectively.}
\label{ee_b} 
\end{figure}

The random experiments are repeated 50 times, and the mean values are compared.
Fig. \ref{ee_b} presents the average energy consumption per user across different deadline ranges.
The number of users is set to $M=10$ and $M=20$, respectively. 
Three different ranges of $\beta$ are evaluated: $[4.5,5.5]$, $[2.0,8.0]$, and $[0.0,10.0]$.
The performance comparison reveals similar trends to those observed in the identical deadline scenario.
The proposed J-DOB consistently outperforms the others when paired with the same outer OG \cite{shi2022multiuser} algorithm, achieving up to a 45.27\% and 44.74\% energy reduction compared to LC for $M=10$ and $M=20$, respectively.

\section{Conclusion}
This paper presents an energy-saving approach for multi-user co-inference under hard latency constraints.
Specifically, we propose J-DOB, an algorithm that near-optimally integrates device-edge joint DVFS, identical offloading, and greedy batching into a unified scheduling framework. 
By efficiently sorting users, calculating edge frequency thresholds, and updating the offloading set, J-DOB maintains low computational complexity.
Experiment results demonstrate that J-DOB significantly reduces energy consumption compared to both local computing and existing methods. Future work will explore online scenarios where precise predictions of future task arrivals are unavailable.
\section*{Acknowledgment}
This work is supported in part by the National Natural Science Foundation of China under Grants 62341108.

\bibliographystyle{ieeetr}  
\bibliography{reference}

\begin{thebibliography}{10}

\bibitem{choi2021lazy}
Y.~Choi, Y.~Kim, and M.~Rhu, ``Lazy batching: {A}n {SLA}-aware batching system for cloud machine learning inference,'' in {\em 2021 IEEE International Symposium on High-Performance Computer Architecture (HPCA)}, pp.~493--506, IEEE, 2021.

\bibitem{you2019large}
Y.~You, J.~Li, S.~Reddi, J.~Hseu, S.~Kumar, S.~Bhojanapalli, X.~Song, J.~Demmel, K.~Keutzer, and C.-J. Hsieh, ``Large batch optimization for deep learning: Training {BERT} in 76 minutes,'' {\em arXiv preprint arXiv:1904.00962}, 2019.

\bibitem{hu2019dynamic}
C.~Hu, W.~Bao, D.~Wang, and F.~Liu, ``Dynamic adaptive {DNN} surgery for inference acceleration on the edge,'' in {\em IEEE INFOCOM 2019-IEEE Conference on Computer Communications}, pp.~1423--1431, IEEE, 2019.

\bibitem{song2021adaptive}
J.~Song, Z.~Liu, X.~Wang, C.~Qiu, and X.~Chen, ``Adaptive and collaborative edge inference in task stream with latency constraint,'' in {\em ICC 2021-IEEE International Conference on Communications}, pp.~1--6, IEEE, 2021.

\bibitem{tang2020joint}
X.~Tang, X.~Chen, L.~Zeng, S.~Yu, and L.~Chen, ``Joint multiuser {DNN} partitioning and computational resource allocation for collaborative edge intelligence,'' {\em IEEE Internet of Things Journal}, vol.~8, pp.~9511--9522, June 2021.

\bibitem{zhang2019mark}
C.~Zhang, M.~Yu, W.~Wang, and F.~Yan, ``{MA}rk: Exploiting cloud services for cost-effective, {SLO}-aware machine learning inference serving,'' in {\em 2019 {USENIX} Annual Technical Conference ({USENIX} {ATC} 19)}, pp.~1049--1062, 2019.

\bibitem{xu2023smdp}
Y.~Xu, J.~Sun, S.~Zhou, and Z.~Niu, ``Smdp-based dynamic batching for efficient inference on gpu-based platforms,'' in {\em ICC 2023-IEEE International Conference on Communications}, pp.~5483--5489, IEEE, 2023.

\bibitem{ali2020batch}
A.~Ali, R.~Pinciroli, F.~Yan, and E.~Smirni, ``Batch: machine learning inference serving on serverless platforms with adaptive batching,'' in {\em SC20: International Conference for High Performance Computing, Networking, Storage and Analysis}, pp.~1--15, IEEE, 2020.

\bibitem{10829781}
Y.~Xu, S.~Zhou, and Z.~Niu, ``Smdp-based dynamic batching for improving responsiveness and energy efficiency of batch services,'' {\em IEEE Transactions on Parallel and Distributed Systems}, pp.~1--16, 2025.

\bibitem{shi2022multiuser}
W.~Shi, S.~Zhou, Z.~Niu, M.~Jiang, and L.~Geng, ``Multiuser co-inference with batch processing capable edge server,'' {\em IEEE Transactions on Wireless Communications}, vol.~22, no.~1, pp.~286--300, 2022.

\bibitem{liu2023resource}
Z.~Liu, Q.~Lan, and K.~Huang, ``Resource allocation for multiuser edge inference with batching and early exiting,'' {\em IEEE Journal on Selected Areas in Communications}, vol.~41, no.~4, pp.~1186--1200, 2023.

\bibitem{cang2024joint}
Y.~Cang, M.~Chen, and K.~Huang, ``Joint batching and scheduling for high-throughput multiuser edge ai with asynchronous task arrivals,'' {\em IEEE Transactions on Wireless Communications}, 2024.

\bibitem{rabaey2003digital}
M.~R. Jan, C.~Anantha, N.~Borivoje, {\em et~al.}, {\em Digital integrated circuits: a design perspective}.
\newblock Pearson, 2003.

\bibitem{zhang2024dvfo}
Z.~Zhang, Y.~Zhao, H.~Li, C.~Lin, and J.~Liu, ``Dvfo: Learning-based dvfs for energy-efficient edge-cloud collaborative inference,'' {\em IEEE Transactions on Mobile Computing}, 2024.

\bibitem{wu2023moc}
Y.~Wu, Y.~Gong, Z.~Zhan, G.~Yuan, Y.~Li, Q.~Wang, C.~Wu, and Y.~Wang, ``Moc: Multi-objective mobile cpu-gpu co-optimization for power-efficient dnn inference,'' in {\em 2023 IEEE/ACM International Conference on Computer Aided Design (ICCAD)}, pp.~1--10, IEEE, 2023.

\bibitem{karzhaubayeva2023cnn}
M.~Karzhaubayeva, A.~Amangeldi, and J.-G. Park, ``Cnn workloads characterization and integrated cpu-gpu dvfs governors on embedded systems,'' {\em IEEE Embedded Systems Letters}, 2023.

\bibitem{nabavinejad2022coordinated}
S.~M. Nabavinejad, S.~Reda, and M.~Ebrahimi, ``Coordinated batching and dvfs for dnn inference on gpu accelerators,'' {\em IEEE Transactions on Parallel and Distributed Systems}, vol.~33, no.~10, pp.~2496--2508, 2022.

\bibitem{hanhirova2018latency}
J.~Hanhirova, T.~K{\"a}m{\"a}r{\"a}inen, S.~Sepp{\"a}l{\"a}, M.~Siekkinen, V.~Hirvisalo, and A.~Yl{\"a}-J{\"a}{\"a}ski, ``Latency and throughput characterization of convolutional neural networks for mobile computer vision,'' in {\em Proceedings of the 9th ACM Multimedia Systems Conference}, pp.~204--215, 2018.

\bibitem{ren2020accelerating}
J.~Ren, G.~Yu, and G.~Ding, ``Accelerating dnn training in wireless federated edge learning systems,'' {\em IEEE Journal on Selected Areas in Communications}, vol.~39, pp.~219--232, Jan. 2021.

\bibitem{sandler2018mobilenetv2}
M.~Sandler, A.~Howard, M.~Zhu, A.~Zhmoginov, and L.-C. Chen, ``Mobilenetv2: Inverted residuals and linear bottlenecks,'' in {\em Proceedings of the IEEE Conference on Computer Vision and Pattern Recognition}, pp.~4510--4520, 2018.

\bibitem{yang2020energy}
Z.~{Yang}, M.~{Chen}, W.~{Saad}, C.~S. {Hong}, and M.~{Shikh-Bahaei}, ``Energy efficient federated learning over wireless communication networks,'' {\em IEEE Transactions on Wireless Communications}, vol.~20, pp.~1935--1949, Mar. 2021.

\bibitem{3090}
NVIDIA, ``{NVIDIA RTX}3090.'' [Online].
\newblock Available: \url{https://www.nvidia.cn/geforce/graphics-cards/30-series/rtx-3090-3090ti/}, (accessed 20-Feb-2025).

\bibitem{wolsey2007mixed}
L.~A. Wolsey, ``Mixed integer programming,'' {\em Wiley Encyclopedia of Computer Science and Engineering}, pp.~1--10, 2007.

\end{thebibliography}

\end{document}